\documentclass[12pt]{article}
\usepackage{pdproc,epsfig} 

\begin{document}

\newcommand{\be}{\begin{eqnarray}}
\newcommand{\ee}{\end{eqnarray}}
\newcommand{\etal}{{\it et al.}}
\def\nue{{\nu_e}}
\def\anue{{\bar\nu_e}}
\def\numu{{\nu_{\mu}}}
\def\anumu{{\bar\nu_{\mu}}}
\def\nutau{{\nu_{\tau}}}
\def\anutau{{\bar\nu_{\tau}}}
\newcommand{\dm}{\mbox{$\Delta m_{21}^2$~}}
\newcommand{\st}{\mbox{$\sin^{2}2\theta$~}}
\newcommand{\br}{\mbox{$^{8}{B}~$}}
\newcommand{\ber}{\mbox{$^{7}{Be}$~}}
\newcommand{\cl}{\mbox{$^{37}{Cl}$~}}
\newcommand{\ga}{\mbox{$^{71}{Ga}$~}}
\newcommand{\chr}{\mbox{$\breve{\rm C}$erenkov~}}
\newcommand{\kl}{\mbox{KamLAND~}}
\newcommand{\bx}{\mbox{Borexino~}}
\newcommand{\thsol}{\mbox{$\theta_{12}$~}}
\def\lsim{\:\raisebox{-0.5ex}{$\stackrel{\textstyle<}{\sim}$}\:}
\def\gsim{\:\raisebox{-0.5ex}{$\stackrel{\textstyle>}{\sim}$}\:}
\def\ltap{\ \raisebox{-.4ex}{\rlap{$\sim$}} \raisebox{.4ex}{$<$}\ }
\def\gtap{\ \raisebox{-.4ex}{\rlap{$\sim$}} \raisebox{.4ex}{$>$}\ }
\newcommand{\ms}{\Delta m^2_{\odot}}
\newcommand{\ma}{\Delta m^2_{\rm atm}}
\newcommand{\ts}{\sin^2 2\theta_{\odot}}
\newcommand{\sss}{\sin^2 \theta_{\odot}}
\newcommand{\sch}{\sin^2 \theta_{13}}

\renewcommand{\thefootnote}{\alph{footnote}}
  
\title{Solar Neutrino Oscillation Parameters in Experiments with Reactor 
Anti-Neutrinos}
 
\author{Sandhya Choubey}

\address{
INFN, Sezione di Trieste, Trieste, Italy\\
Scuola Internazionale Superiore di Studi Avanzati,
I-34014,
Trieste, Italy\\
 {\rm E-mail: sandhya@he.sissa.it}}

\abstract{
We review the current status of the solar neutrino oscillation 
parameters. We discuss the conditions under which measurements 
from future solar neutrino experiments would determine the oscillation 
parameters precisely. Finally we expound the
potential of long baseline 
reactor anti-neutrino experiments in measuring the 
solar neutrino oscillation parameters.
}
   
\normalsize\baselineskip=15pt

\section{Introduction}
Recent data  
on charged current (CC) and neutral current (NC) 
break-up of deuterons by the $^8B$ solar neutrinos 
from the Sudbury Neutrino Observatory (SNO) \cite{sno1,sno2,sno3},
have confirmed the 
solar neutrino deficit problem, first observed in the 
pioneering experiment at Homestake (Cl), 
and later corroborated by the 
observations in the SAGE, GALLEX, GNO (Ga) and Kamiokande and 
Super-Kamiokande (SK) experiments \cite{sol}. 
Spearheaded by the CC to NC ratio observed in SNO, the global 
solar neutrino data collected over the last few decades, 
established the Large Mixing Angle (LMA) solution as the most favored 
solution to the solar neutrino deficit problem 
\cite{sno3,snocc,snonc,salt,saltother}. 
The other oscillation solutions such as SMA, LOW, QVO and VO are strongly 
disfavored, although alternative 
mechanisms involving flavor changing neutral currents (FCNC) 
or transition magnetic moment (RSFP) \cite{rsfp} would still be allowed 
by the global solar neutrino data.
The KamLAND reactor anti-neutrino experiment has 
observed flavor oscillations of $\anue$ and 
therefore under the plausible assumption of CPT invariance has  
independently given a conclusive evidence in favor of the LMA 
solution \cite{kamland}. The global data, including the KamLAND 
and the solar neutrino results pick $\ms \equiv \dm 
\approx 7\times 10^{-5}$ 
eV$^2$ and $\sss \equiv \sin^2\theta_{12} 
\approx 0.3$ as the best-fit solution
\cite{kldata}. KamLAND 
virtually ``rules out'' all the alternative solution 
to the solar neutrino problem and relegates them to 
play at best a sub-dominant role in the deficit of solar neutrinos.

The observed depletion of the atmospheric muon neutrino flux
in SK, and in particular the Zenith angle dependence of this 
observed deficit, have given strong evidence for the existence 
of oscillations of atmospheric muon neutrinos. The SK 
atmospheric data is best explained
in terms of dominant $\nu_{\mu} \rightarrow \nu_{\tau}$ 
($\bar{\nu}_{\mu}\rightarrow \bar{\nu}_{\tau}$) oscillations
with  maximal mixing and $\Delta m_{\rm atm}^2 \approx 2\times 10^{-3}$ 
eV$^2$ \cite{SKatmo03}.

The two sectors, solar and atmospheric, are related by the mixing 
angle $\theta_{13}$ which is currently bound by the CHOOZ and Palo-Verde 
\cite{CHOOZPV}
reactor data as $\sin^2\theta_{13} \ltap 0.1$, dependending on the 
``true'' value of  $\Delta m_{\rm atm}^2$.

\section{Status of the Solar Neutrino Oscillation Parameters}

\begin{figure}
\epsfig{file=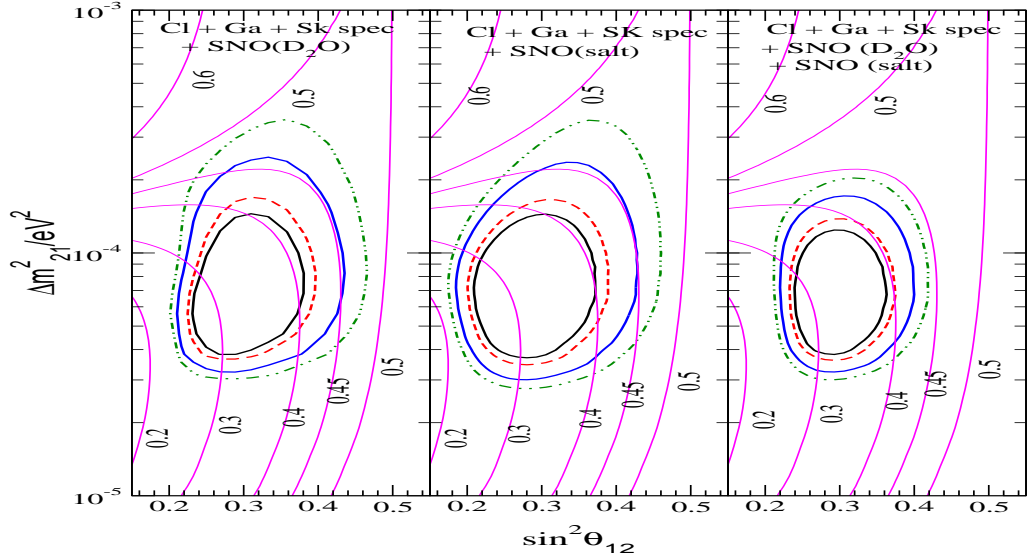,width=6in,height=4.0in}
\caption{The 90\%, 95\%, 99\% and 99.73\% C.L. 
allowed regions in the $\ms-\sss$ plane from global 
$\chi^2$-analysis of the data from solar neutrino experiments.
We use the $\Delta \chi^2$ values 
corresponding to a two parameter fit
to plot the C.L. contours.
Also shown are the lines of constant CC/NC event rate 
ratio $R_{CC/NC}$.
}
\label{sol}
\end{figure}

In figure \ref{sol} 
we show the areas of the solar neutrino oscillation 
parameter space, which are allowed by the global solar neutrino data. 
In particular, this figure shows the impact of the salt phase of 
the SNO results \cite{sno3}. In the left-hand panel of the figure 
\ref{sol}, we show the allowed areas obtained when global solar data 
includes only the $D_2O$ phase of SNO. In the middle panel we show the 
corresponding contours when global solar data 
includes only the salt phase of SNO. Finally, in the 
right-hand panel we give the allowed areas obtained 
from the global solar neutrino data including SNO results from 
both the $D_2O$ and the salt phases.
Also superimposed on the figures are the constant lines of the ratio of 
the CC to NC rates, $R_{CC/NC}$ in SNO \cite{maris}.
We note that smaller values 
of $R_{CC/NC}$ roughly trace smaller values of $\ms$ and $\sss$. 
In particular, since the value of this ratio has reduced from 
$R_{CC/NC} = 0.346$ as in the $D_2O$ phase to $R_{CC/NC} = 0.306$ as 
in the salt phase, comparison of the left and the middle panels of 
figure \ref{sol} reflect the fact that for the salt phase the allowed 
areas have shifted to lower values of $\sss$ (the impact on $\ms$ 
is not seen to be very significant). 
The third panel shows the allowed areas after including data from 
both the $D_2O$ and salt phase of SNO. The combined SNO data 
has a much larger statistical power and this results in tighter 
constraints both on the upper bound on $\ms$ and upper bound on $\sss$. 
SNO also disfavors maximal mixing at the $5.4\sigma$ level.

\begin{figure}
\epsfig{file=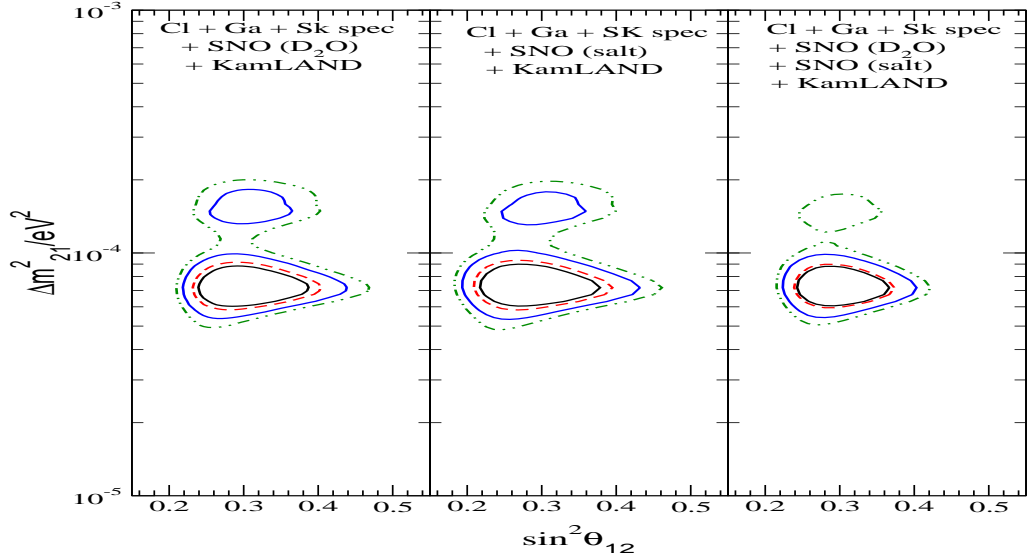,width=6in,height=4.0in}
\caption{The 90\%, 95\%, 99\% and 99.73\% C.L.
allowed regions in the $\ms-\sss$ plane from global 
$\chi^2$-analysis of solar and \kl data.
We use the $\Delta \chi^2$ values corresponding to a
2 parameter fit 
to plot the C.L. contours.}
\label{solkl}
\end{figure}

\begin{table}
\begin{center}
\begin{tabular}{c|cc|cc}
\hline\hline
Data & \multicolumn{2}{c|} {best-fit parameters}  & \multicolumn{2} {c}
{99\% C.L. allowed range}\\
\cline{2-5}
set used & \dm/($10^{-5}$eV$^2$) & $\sss$ & \dm/($10^{-5}$eV$^2$) & $\sss$\\
\hline \hline
Cl+Ga+SK+$D_2O$& $6.06$ & 0.29 & $3.2-24.5$ &  $0.21 - 0.44$ \\
Cl+Ga+SK+salt& $6.08$ & 0.28 & $3.0-23.7$ &  $0.19 - 0.43$ \\
Cl+Ga+SK+$D_2O$+salt& $6.06$ & 0.29 & $3.2-17.2$ &  $0.22 - 0.40$ \\
Cl+Ga+SK+$D_2O$+KL& $7.17$ & 0.3 & $5.3-9.9$ &  $0.22 - 0.44$ \\
Cl+Ga+SK+$D_2O$+salt+KL& $7.17$ & 0.3 & $5.3-9.8$ &  $0.22 - 0.40$ \\
\hline \hline
\end{tabular}
\label{tab1}
\caption{
The best-fit values of the solar neutrino oscillation 
parameters, obtained using different combinations 
of data sets. Shown also are the 99\% C.L. (corresponding to $\Delta \chi^2$ 
for a 2 parameter fit) allowed ranges of the parameters from 
the different analyses.
}
\end{center}
\end{table}

We next include the 162 ton-year first results from 
\kl \cite{kamland} in the analysis and present 
the corresponding allowed areas obtained from the combined \kl + solar 
analysis in figure \ref{solkl}. 
Again we present our results separately for three cases for SNO: 
with the global solar data including only the 
$D_2O$ phase (left-hand panel), with the global solar data including 
only the salt phase (middle panel) 
and with the global solar data including 
the two phases combined (right-hand panel). 
We note the shift in the allowed zones to smaller values of $\sss$ 
for the salt phase panel in figure \ref{solkl}. This is due 
to the change in the value of $R_{CC/NC}$ for SNO, as discussed above.
The main impact of the 
\kl data is to split the LMA zone into two allowed sub-zones, 
which we will call the low-LMA and the high-LMA, with best-fit 
$\ms$ around $7.2\times 10^{-5}$ eV$^2$ and $1.5\times 10^{-4}$ eV$^2$
respectively. The best-fit value of $\sss=0.3$ for both the solutions.
We note that the high-LMA solution is allowed at the 99\% C.L.
when the two phases of the SNO data are 
included separately. However the combined SNO data along with the 
other solar neutrino data and the \kl results, allow 
the high-LMA 
only at 99.13\% C.L. (2.63$\sigma$)
with respect to the 
global $\chi^2_{min}$ obtained in the low-LMA region.

In Table 1 we show the best-fit points and the 99\% C.L. 
allowed range of parameter values for the different combination 
of data sets. We note that while the inclusion of 
\kl data severely restricts the range 
of allowed values for the mass difference $\ms$, the range of allowed 
values for $\sss$ remains virtually the same as that allowed by the 
global solar neutrino data alone.

\section{Potential of the Future Solar Neutrino Experiments}

With LMA confirmed as the solution to the solar neutrino 
problem, the stage is set for the herald of the era of precision 
measurement in the field of neutrino physics. We would first want to 
glean into the immediate future and see how much light
the next generation results from solar neutrino experiments
could shed into our understanding of the solar neutrino oscillation 
parameters.

After having completed the very successful salt phase of their 
experiment, the SNO collaboration will use Helium proportional 
counters to observe directly the neutrons released in the neutral 
current break-up of deuteron. This phase 3 result from SNO will give 
a totally uncorrelated and clean signal for the observed
CC and NC event rates. The phase 3 results will also have higher 
statistics and therefore will be further constraining in $\ms$ and 
$\sss$. In the near future, SNO is expected to provide data on the 
day/night spectrum, which could be used in a statistical 
analysis to further constrain 
the solar neutrino oscillation 
parameters. One of the related observables
is the day-night asymmetry:
\be
A_{DN} = 2\frac{N-D}{N+D}.
\ee
The predicted $A_{DN}$ in SNO,
for the current best-fit values of the parameters  
in the low-LMA region, as well as the corresponding
$3\sigma$ range, are given by 
\be
A_{DN}^{SNO} = 0.04,~~3\sigma~ {\rm range}:~0.02-0.07,~~~{\rm low-LMA}, 
\ee
For the barely allowed high-LMA solution we get:  
\be
A_{DN}^{SNO} = 0.01,~~3\sigma~ {\rm range}:~0.007-0.02,~~~{\rm high-LMA}. 
\ee

\begin{figure}
\epsfig{file=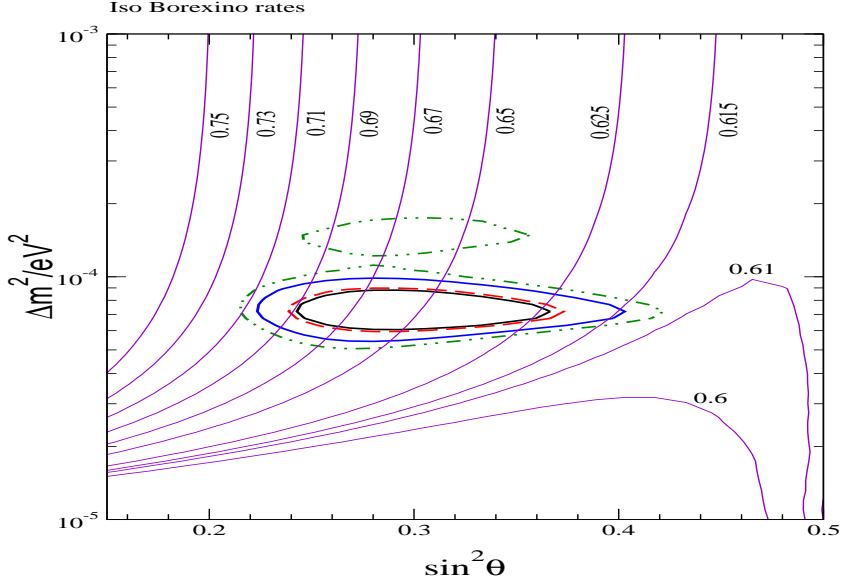,width=5in,height=4.0in}
\vskip -1cm
\caption{The isorate lines for the Borexino detector in the
$\ms-\sss$ plane.
Also shown are the C.L. contours from the global analysis 
of the solar and the \kl data. 
}
\label{be}
\end{figure}

\begin{figure}
\epsfig{file=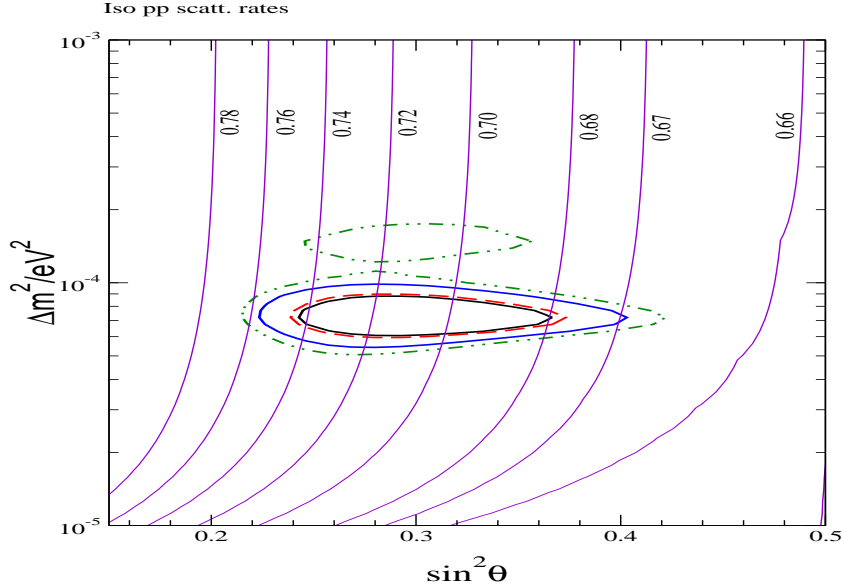,width=5in,height=4.0in}
\vskip -1cm
\caption{The isorate lines for a generic $pp$ - e scattering 
experiment in the 
$\ms-\sss$ plane.
Also shown are the C.L. contours from the global analysis 
of the solar and the \kl data.
}
\label{pp}
\end{figure}

The potential of Borexino \cite{borex} and any
generic 
electron scattering experiment for the low energy $pp$ neutrinos -- the 
LowNu experiments \cite{lownu} -- in constraining the 
mass and mixing parameters have been studied most recently in 
\cite{th12,Bahcall:2003ce}.
For the current range of allowed parameter values, 
we find the predicted rates
for Borexino and LowNu experiments to be 
\be
R_{Be} &=& 0.65,~~(3\sigma~ {\rm range} \equiv 0.61-0.71);~~~{\rm low-LMA}\\
R_{pp} &=& 0.71,~~(3\sigma ~{\rm range} \equiv 0.67-0.76);~~~{\rm low-LMA}
\ee
Figures \ref{be} and \ref{pp} show the iso-rate contours for 
the observed rates in Borexino and a generic LowNu experiment.
Also superimposed on the figures are the current allowed zone 
from the global solar and reactor data.
The figure \ref{be} shows that there is {\it almost} no $\ms$ dependence 
for the observed rate in Borexino over most of the allowed range of 
the parameter space. Thus we would not 
expect the range of $\ms$ to improve much with Borexino. The Borexino  
rate is seen to have some dependence on the value of $\sss$. 
Thus the range of $\sss$ could be improved upon if Borexino would be 
successful in measuring the solar $^7Be$ rate with a $1\sigma$ 
experimental error of 
less than about $2-3\%$ \cite{th12}. Indeed it was shown in 
\cite{Bahcall:2003ce} as well that even with 5\% error in 
the observed rate in Borexino, the range of allowed values of $\ms$ and 
$\sss$ do not change much.

The figure \ref{pp} shows that the iso-rates for 
a generic $pp$ neutrino-electron scattering experiment are also 
nearly independent of $\ms$ but carry a fair degree of dependence 
on the value of $\sss$. The advantage of these kind of experiments 
is that the $pp$ flux is theoretically known to within 1\% accuracy 
\cite{bbp00}. However one would still need an experimental error 
of less that a few percent to be able to constrain $\sss$ any further
\cite{Bahcall:2003ce}.

\section{Potential of KamLAND for Precision Measurement}

\begin{figure}[t]
\centerline{\psfig{figure=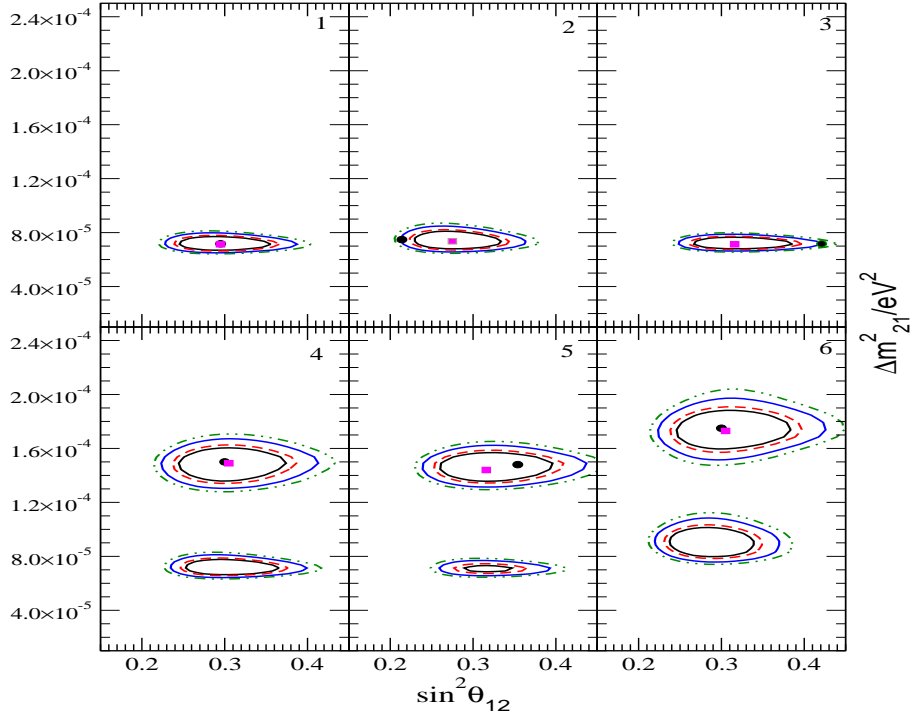,height=4.8in,width=5.in}}
\caption{
The 90\%, 95\%, 99\% and 99.73\% C.L. allowed regions obtained from a 
combined analysis using the global solar neutrino data and a 
1.0 kTy simulated KamLAND data. The points in the parameter 
space, for which the 1.0 kTy \kl data has been simulated, are shown 
by the black dots; they have been chosen to lie 
within the current $3\sigma$ allowed regions.
The best-fit point of the combined analysis 
are shown as red ``boxes''. 
Also shown superimposed are current global the 99.73\% C.L.allowed areas.
}
\label{kl1yr}
\end{figure}
\begin{table}
\begin{center}
\begin{tabular}{ccccc}
\hline
$ \!\!\!\!$Data$ \!\!\!\!$& $ \!\!\!\!$99\% CL $ \!\!\!\!$ 
&$ \!\!\!\!$$ \!\!\!\!$99\% CL $ \!\!\!\!$ &$ \!\!\!\!$ 99\% CL 
$ \!\!\!\!$& $ \!\!\!\!$99\% CL $ \!\!\!\!$\cr
$ \!\!\!\!$set$\!\!\!\!$ & range of$ \!\!\!\!$ 
& spread $ \!\!\!\!$&range $ \!\!\!\!$ 
& $ \!\!\!\!$spread  $ \!\!\!\!$
\cr
$ \!\!\!\!$used$ \!\!\!\!$& $\ms\times$ $ \!\!\!\!$& of $ \!\!\!\!$& 
of$ \!\!\!\!$ &$ \!\!\!\!$ in  $ \!\!\!\!$\cr
$ \!\!\!\!$
$\!\!\!\!$& 10$^{-5}$eV$^2$ & \dm $ \!\!\!\!$& 
$\sin^2\theta_{\odot}$ $ \!\!\!\!$
&$ \!\!\!\!$ $\sin^2\theta_{\odot}$ $ \!\!\!\!$\cr
\hline
$ \!\!\!\!$only sol$\!\!\!\!$ & 3.2 - 17.0 $ \!\!\!\!$ 
& 68\% $ \!\!\!\!$& $0.22-0.40$ $ \!\!\!\!$&$ \!\!\!\!$ 29\%$ \!\!\!\!$ \cr
$ \!\!\!\!$sol+162 Ty$ \!\!\!\!$&  5.3 - 9.8 $ \!\!\!\!$
& $ \!\!\!\!$30\% $ \!\!\!\!$
& $0.22-0.40$$ \!\!\!\!$ &$ \!\!\!\!$ 29\% $ \!\!\!\!$ \cr
$ \!\!\!\!$sol+1 kTy$\!\!\!\!$ & 6.5 - 8.0 $ \!\!\!\!$
& 10\%$ \!\!\!\!$ & 
$0.23-0.39$ $ \!\!\!\!$&$ \!\!\!\!$ 26\% $ \!\!\!\!$\cr
$ \!\!\!\!$sol+3 kTy $\!\!\!\!$& 6.8 - 7.6 $ \!\!\!\!$
& 6\% $ \!\!\!\!$& $0.24-0.37$$ \!\!\!\!$ &$ \!\!\!\!$ 21\% $ \!\!\!\!$\cr
\hline
\end{tabular}
\caption
{The range of parameter values allowed at 99\% C.L.
and the corresponding spread. 
}
\label{klbounds}
\end{center}
\end{table}

The \kl experiment with their first results have already shown 
remarkable promise to measure precisely the  
solar neutrino oscillation parameters. In figure \ref{kl1yr} we show the 
allowed areas obtained from a combined analysis performed with the 
current global solar neutrino data and a projected 1 kTy \kl data 
simulated at various points shown in the figure by black dots. 
We note that the precision on the allowed value of $\ms$ improves, 
however for $\sss$ there seem to little improvement. Also note that 
if the future \kl spectral data corresponds to a point in the 
high-LMA zone, then the ambiguity between the high-LMA and 
low-LMA solution would again get enhanced to the 90\% C.L. due 
to the conflicting trends between the solar and \kl data.

We present in 
Table 2 the values of $\ms$ and $\sss$
allowed at 99\% C.L. by the existing and 
prospective \kl data, 
and the corresponding uncertainty (``spread'') defined as 
\cite{th12,th12hlma},
\be
{\rm spread} = \frac{p_{\rm max}-p_{\rm min}}{p_{\rm max}+p_{\rm min}}
\ee
where $p_{\rm max(min)}$ is the largest(smallest) allowed value 
of the given parameter.
The uncertainty in $\ms$, determined using 
only the $\nu_{\odot}$ data, reduces from 68\% 
to 30\% after the inclusion of the first \kl data 
in the analysis, while that in $\sss$ does not change,
remaining rather large - 29\%.
The uncertainty in $\ms$ would further diminish 
to 10\% (6\%) after 1 kTy (3 kTy) data from KamLAND.
However, there is little 
improvement in the precision on the value of 
$\sss$ with the increase of 
\kl statistics \cite{th12}.

\section{The importance of the SPMIN}

  The $\bar{\nu}_e$ survival probability, $P_{ee}$, in 
the reactor experiments of interest,  
depends on $\ms$, $\sss$, $\ma$,  
the angle $\theta_{13}$
limited by the CHOOZ and Palo Verde experiments, 
and on the type of neutrino mass hierarchy 
\cite{th12hlma,SPMP02}.
The potential sensitivity of a reactor experiment
to each of these parameters 
depends crucially on the baseline of the experiment.
Experiments with a baseline 
$L \sim (1 - 2)$ km can be used
to get information on $\sch$, since
over these distances
oscillations 
induced by $\ma$ are mainly operative
and $\sin^22\theta_{13}$ determines 
their amplitude. For baselines
$L \gtap 50$ km,
the $\bar{\nu}_e$ oscillations
due to $\ma$ 
average out and we have,
$P_{ee} \approx [1 - \sin^2 2\theta_{\odot}
\sin^2(\ms L/4E)] \cos^4\theta_{13}$.
Therefore long baseline reactor experiments can measure 
$\ms$ and $\sss$.

In the absence of oscillations, 
the maximal contribution 
to the signal in a reactor experiment
comes for $\bar{\nu}_e$ energy $E \sim 3.6$ MeV.
For a fixed $\ms$, maximal sensitivity to 
$\sss$ can be achieved 
if for $E \sim 3.6$ MeV,
$L$ is ``tuned'' to
a $\bar{\nu}_e$ survival probability
minimum (SPMIN), i.e., if 
$\sin^2(\dm L/4E) \approx 1$.
The corresponding
$P_{ee} \approx 1 - \sin^2 2\theta_{\odot}$,
and thus very sensitive to 
the value of $\sin^2 2\theta_{\odot}$. 
If in contrast, $L$ is such that 
$\sin^2(\dm L/4E) = \epsilon \approx 0$,
$P_{ee}$ would have a maximum
(SPMAX): $P_{ee} \approx 1 - 
\epsilon \sin^2 2\theta_{\odot} \approx 1$.
In this case the sensitivity
to $\sin^2 2\theta_{\odot}$
is worse than in the preceding 
one. The positions of the extrema 
in both cases are highly sensitive to the 
value of $\ms$. We note that for \kl the dominant reactor 
anti-neutrino flux coming from the Kashiwazaki reactor power 
complex corresponds to a SPMAX for the low-LMA and the high-LMA solutions. 
This is why the $\sss$ sensitivity is 
not good in KamLAND.

\section{Impact of new reactors on \kl}

\begin{figure}
\begin{center}
\vspace{0.3cm} \epsfxsize = 9.8cm \epsffile{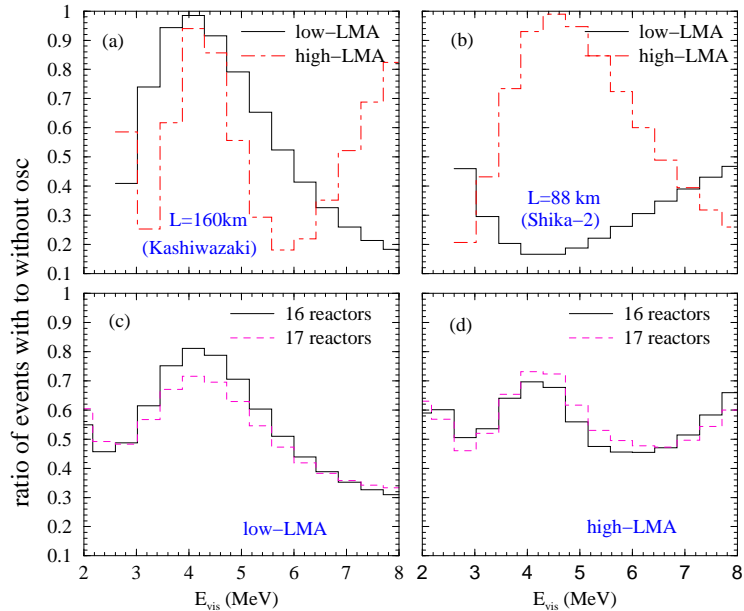}
\leavevmode
\caption{\label{fig1}
The spectral distortion expected at KamLAND.
The upper panels show the contribution from the individual 
fluxes from Kashiwazaki (panel (a)) and Shika-2 (panel (b)).
The lower panels show the cumulative resultant spectral 
distortion for the low-LMA (panel (c)) and high-LMA (panel (d)) 
solutions. 
}
\end{center}
\end{figure}

\begin{figure}
\begin{center}
\vspace{0.3cm} \epsfxsize = 10.5cm \epsffile{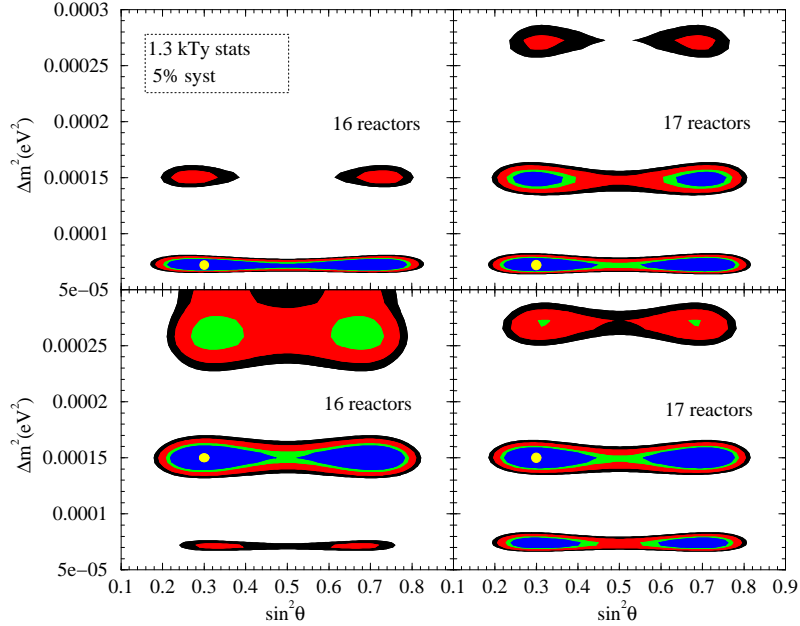}
\leavevmode
\caption{\label{fig2}
Prospective 90\%, 95\%, 99\% and 99.73\% C.L. 
contours in the $\ms - \sss$ plane,
which would be obtained using the 
\kl data corresponding to 1.3 kTy. 
The points at which 
the spectrum was simulated are shown by yellow circles.
}
\end{center}
\end{figure}

A new reactor power complex called ``Shika-2'', with a thermal power 
of about 4 GW and close to the old Shika site, at a distance of 
about 88 km from KamLAND, 
is expected to start operations from March 2006 
onward. In Fig. \ref{fig1}, the upper panels show the  
spectrum of the  
observed to expected positron events ratio in KamLAND, induced by the 
``individual'' reactor fluxes from the Kashiwazaki (panel (a))
and the Shika-2 (panel (b)) 
reactor complexes. 
The spectra corresponding to both low-LMA and 
high-LMA solutions are displayed. We note that for 
the Kashiwazaki flux, the events spectra for both low- and 
high-LMA correspond to roughly a SPMAX. Since the 
$\anue$ flux from Kashiwazaki dominates the total observed flux 
at KamLAND, 
the fact that for both low- and high-LMA solutions the 
Kashiwazaki flux 
produces a similar 
spectrum at KamLAND, results in the two degenerate solutions 
allowed by the current data.

For Shika-2, the 
low-LMA solution produces a SPMIN in the resultant spectrum 
at the detector, while 
the high-LMA solution 
produces a SPMAX. 
Therefore, with the inclusion of the Shika-2 flux, we could expect 
that the ({\it i}) ability of \kl to discriminate between the 
low- and the high-LMA solutions should improve and 
({\it ii}) in the case of low-LMA as the true solution, 
sensitivity of \kl to measure $\sss$ should improve.

The bottom panels in Fig. \ref{fig1} show the positron spectrum 
corresponding to the cumulative flux seen at \kl from all the 
reactors combined. Panel (c) shows the spectrum corresponding to 
the low-LMA solution while panel (d) shows the spectrum expected 
for the high-LMA solution. The solid line in both panels correspond 
to the cumulative spectrum from the current 16 main reactors 
operating around KamLAND, while the dashed lines show the case when 
the Shika-2 reactor also starts operation, along with the 16 already 
existing reactor facilities. Clearly we see that even 
after the starting of the 
Shika-2 reactor, the resultant spectrum is a SPMAX for the 
low-LMA solution. This happens because the Kashiwazaki 
power plant even though farther from \kl, is much more powerful 
than Shika-2. The effective $\anue$ of Kashiwazki at \kl is 
$\approx 7.3 \mu W/cm^2$, which should be compared with 
Shika-2 effective flux of only $\approx 4.1 \mu W/cm^2$. 
Further, a comparison of the 
panels (c) and (d) show that the impact of the Shika-2 flux is 
actually to reduce the difference between the spectral 
distortion produced in the case of the low- and high-LMA solutions.
Therefore instead of improving the sensitivity of \kl to 
distinguish between the two solutions, the effect of turning on the 
Shika-2 $\anue$ flux could further decrease it \cite{shika}.

In Fig. \ref{fig2} we show the projected allowed areas obtained 
by analysing a prospective 1.3 kTy \kl data with either the 
current 16 reactor fluxes (left-hand panels) or with the 
17 reactor setup, including the Shika-2 flux (right-hand panels).
The upper (lower) panels for both the setups 
are for the low-LMA (high-LMA) as 
the true solution: which means that we simulate the prospective \kl 
spectrum at the low-LMA (high-LMA) best-fit. We note that with 
the addition of the Shika-2 flux into the \kl spectrum, the 
spurious high-LMA (low-LMA) solution gets allowed at even the 
90\% C.L., showing a clear deteoration in the experiment's 
ability to pick the right solution. The sensitivity to $\sss$ does 
not improve either \cite{shika}.

\section{Optimizing the Reactor Anti-Neutrino Experiment for $\ms$ and 
$\sss$}

\begin{figure}[t]
\centerline{\psfig{figure=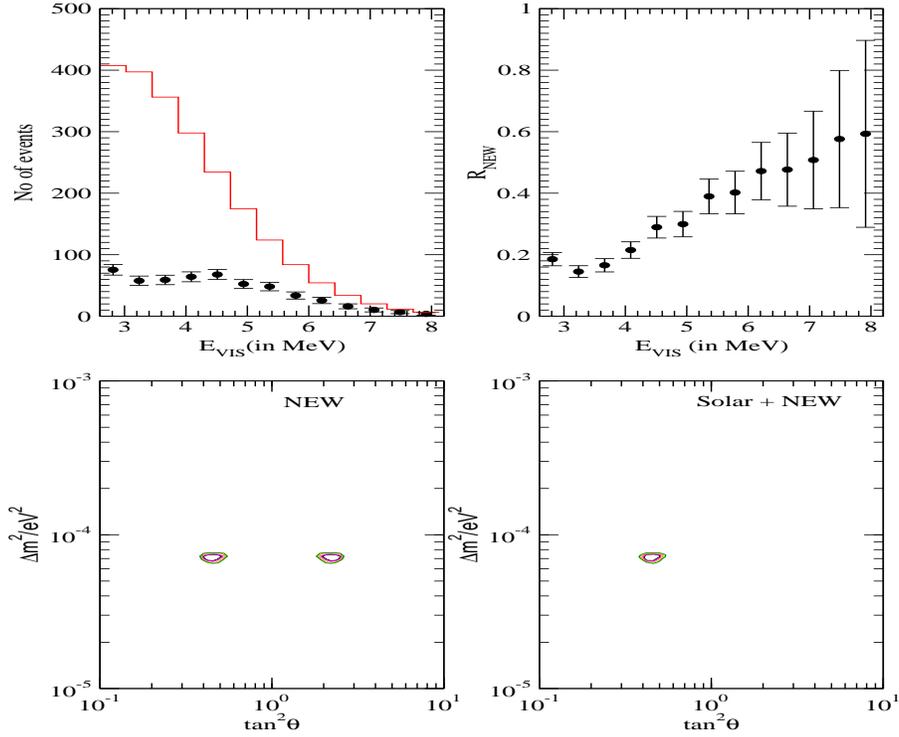,height=4.in,width=5.in}}
\caption{The simulated 3 kTy spectrum data at the low-LMA 
best-fit point and the allowed areas in the $\ms-\tan^2\thsol$ 
parameter space
for a 24 GWatt reactor experiment with 
a baseline of 70 km. The top-left panel gives the 
simulated spectrum data and the expected events, shown by the histograms.
The top-right panel shows $R_{NEW}$, the corresponding ratio  
of the ``data'' to expected 
events as a function of the visible energy. The bottom-left panel 
gives the allowed areas obtained using just the new reactor experiment.
The bottom-right panel presents the allowed areas from the 
combined solar and new reactor experiment data.
}
\label{new}
\end{figure}

For $\ms$ in the low-LMA region, we expect to find a 
minimum in the survival probability (SPMIN) 
when $L\sim 70$ km.
This value of $L$ is therefore
best suited for measuring $\theta_{\odot}$
if $\ms$ lies in the low-LMA region \cite{th12}.
For a reactor complex having a power of 24.6 GW 
(e.g., Kashiwazaki) and data of 3 kTy from a 
\kl-like detector at $L \sim 70$ km, 
$\sss$ can be determined 
with a $\sim$10\% uncertainty \cite{th12}. In figure \ref{new} we 
show the allowed areas in the parameter space we would expect 
after 3 kTy of data from this experiment 
if low-LMA was the true solution to the solar neutrino problem.

\section{Conclusions}

In conclusion, future solar neutrino data can lead to 
precise measurement of the neutrino oscillation parameters 
only if they can reduce the experimental uncertainties.
The KamLAND experiment can measure 
the solar mass squared difference very precisely but not the 
mixing angle. If low-LMA is confirmed by the next results from 
KamLAND, a reactor experiment with a baseline of 70 km should be
ideal to measure the solar neutrino mixing angle. 
A new reactor power plant, Shika-2, 
is expected to start operations in Japan in March 
2006. It will be located at $L \sim 88$ km from \kl. 
This baseline is close 
to the ``ideal'' one of $L \sim 70$ km. 
However 
due to averaging effects of 
the anti-neutrino fluxes from the Kashiwazaki and 
Shika-2 reactors, the sensitivity of 
\kl to $\sss$ would not improve,
while its sensitivity to \dm would diminish.

\section{Acknowledgments}
It is a pleasure to thank Milla Baldo Ceolin for her hospitality at 
the very stimulating NO-VE conference in Venice.
The results summarized in the present note 
were obtained in collaboration with A. Bandyopadhyay, S. Goswami, 
S.T. Petcov and D.P. Roy. 

\end{document}